\documentclass[prd,aps,12pt]{revtex4}
\usepackage{graphicx}

\begin{document}

\thispagestyle{empty}
\begin{center}
\hfill{\small {hep-ph/0210064}}\\
\hfill{\small {BU-HEPP-02/10}}\\
\hfill{\small {TRI-PP-02-15}}\\[2cm]
{\Large The nucleon's strange electromagnetic and scalar matrix elements}
\vspace{5mm}

Randy Lewis$^a$, W. Wilcox$^b$ and R. M. Woloshyn$^c$
\vspace{5mm}

$^a$Department of Physics, University of Regina, Regina, SK, S4S 0A2, Canada \\
$^b$Department of Physics, Baylor University, Waco, TX, 76798-7316, U.S.A. \\
$^c$TRIUMF, 4004 Wesbrook Mall, Vancouver, BC, V6T 2A3, Canada
\vspace{5mm}

(2 October 2002)
\vspace{15mm}

{\bf Abstract}
\end{center}

Quenched lattice QCD simulations and quenched chiral perturbation theory are
used together for this study of strangeness in the nucleon.  Dependences
of the matrix elements on strange quark mass, valence quark mass
and momentum transfer are discussed in both the lattice and chiral
frameworks.  The combined results of this study are in good agreement with
existing experimental data and predictions are made for upcoming experiments.
Possible future refinements of the theoretical method are suggested.
\newpage

\section{Introduction}

The effects of virtual strange quarks on the properties of a single
nucleon represent basic information about QCD and the strong interaction.
Hence, there is presently a great deal of enthusiasm for studies of
the nucleon's strangeness electric and magnetic form factors.
Recent experiments have produced two measurements\cite{SAMPLE,HAPPEX}
and ongoing efforts are expected to provide more results soon\cite{A4G0H2}.

First principles calculation from QCD requires the use of lattice field
theory techniques, and a number of explorations have been carried out
by various authors\cite{LeiTho,GMKentucky,Wilcox,emi,lat02}.
The presence of the disconnected
strange quark loop and the smallness of the resulting strangeness
form factors cause lattice simulations to be expensive and the extraction
of meaningful results to be difficult\cite{emi}.

Chiral perturbation theory (ChPT) can play a valuable complementary role
alongside lattice QCD.
ChPT is QCD's low-energy effective theory written in terms of the physical
hadrons rather than quarks and gluons, and it contains low energy constants
(LEC's) whose numerical values should be determined from lattice QCD or
directly from experiment.  Quenched SU(3) ChPT\cite{qchpt,LabSha} corresponds to
quenched QCD with three active quark flavours --- up, down and strange ---
and it produces analytic expressions for the
strangeness form factors that explicitly display their dependences on
the strange quark mass, valence quark mass and momentum transfer.
It is clearly advantageous to relegate as much of the calculation as possible
to ChPT so that valuable computer time can be spent on the physics that
ChPT cannot predict.  In other words, one need only extract the required
LEC's from lattice QCD simulations, and then the strangeness form factors can
be studied directly in quenched SU(3) ChPT.

On the other hand, the strangeness form factors can in principle
be measured in lattice QCD
simulations with minimal recourse to ChPT: the strange quark
mass and the momentum transfer can be fixed to their physical values in
a lattice simulation and
then ChPT is only needed for extrapolation of the valence quark mass.
This extrapolation can be performed with quenched SU(2) ChPT rather than SU(3),
thereby providing the benefit of a more rapid convergence for the
chiral expansion since it no longer requires expansion in powers of the
strange quark mass\cite{ChenSav}.

In the present work, we report the results of high-statistics lattice QCD
simulations for the strangeness electric and magnetic form factors together
with the strangeness scalar density.  A number of different analysis 
methods are employed
and found to give consistent results.  Two strange quark masses, three
valence quark masses and five momentum transfer values are studied.
We also present the analytic quenched SU(3) ChPT formulae for the three
strangeness matrix elements of interest and apply them to our lattice QCD data.
The alternative of using quenched SU(2) ChPT is briefly discussed as well.
Finally, we compare our results to the existing experimental measurements,
make predictions relevant to upcoming experiments, and suggest directions
for future theoretical work.

Our main conclusions are that the raw lattice results
for the strangeness electric and magnetic form factors 
(before any use of ChPT) are very small, 
that ChPT-based extrapolation to the physical up and down quark mass region
does not substantially change this,
and that the lattice QCD 
predictions are therefore consistent with existing experimental results.

\section{Numerical simulations}

The gauge field configurations used in this study were generated from
the Wilson gauge action at $\beta=6$ on $20^3\times32$ lattices,
corresponding to a lattice spacing of
\begin{equation}\label{adef}
a = 0.1011(7) \text{~fm}
\end{equation}
as obtained by the authors of Ref.~\cite{Gockeler} from a physical
string tension of $\sqrt{K}=427$ MeV.  Actually the lattice spacing is not
uniquely determined in the quenched approximation, and the authors of
Ref.~\cite{GMKentucky} used the physical nucleon mass to arrive at
$a=0.115(6)$ fm.  Our full ensemble
of 2000 configurations was produced from various independently thermalized
Markov chains.  Within each chain, either 2000 or 5000 triple-step
heatbath updates (i.e. applied to three SU(2) subgroups) were executed between
saved configurations.

The Wilson fermion action was used to obtain three valence quark propagators
per configuration, having $\kappa_v=0.152$, 0.153 and 0.154.  These correspond
to pion masses of 
\begin{equation}
am_\pi=0.4772(9)^{+9}_{-2} \text{\cite{QCDPAX}},~~~
0.4237(8)\text{\cite{Gockeler}} \text{~~~and~~~} 0.364(1)\text{\cite{Allton}}
\end{equation}
respectively.  The valence
quarks in our simulations have Dirichlet time boundaries; the source is
four timesteps away from the boundary.  On our $20^3\times32$ lattices,
the five smallest momentum squared values are
\begin{equation}
a^2\vec{q}^{\,2} = n(\pi/10)^2,~~~n=0,1,2,3,4.
\end{equation}
Tabulated in Table~\ref{tab:2point} are the energies of a nucleon having
degenerate quarks and each of these momentum squared values.

Strangeness matrix elements are calculated using standard methods.
This involves a three-point function in which a strange-quark loop is 
correlated with the nucleon propagator. 
It is prohibitively expensive to compute the strange quark loop exactly at
every lattice site, so we employ a stochastic estimator with real $Z_2$
noise\cite{DongLiu}.  To reduce the variance, the first
four terms in the $\kappa_l$ expansion ($\kappa_l$ denoting the loop quark's
hopping parameter) of the quark matrix were subtracted
for the strangeness electric and magnetic form factors, and the
first five terms were subtracted for the scalar density\cite{subtraction}.
This stochastic estimation method is unbiased with any number of noises,
and the statistical uncertainties associated with this noisy estimator
decrease as one increases the number of
noises and/or the number of gauge field configurations.

\begin{figure}[thb]
\includegraphics[width=150mm]{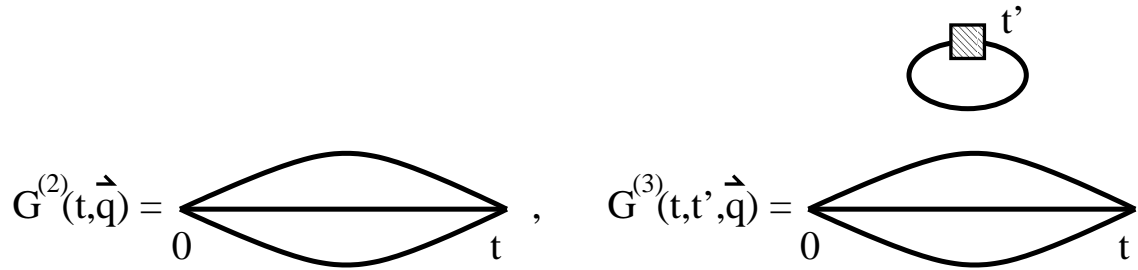}
\caption{Two-point and three-point correlators that appear in $R_X$ of
         Eq.~\protect(\ref{RX}).  Each solid line represents a quark
         propagator, and the shaded box denotes a current
         insertion.}\label{fig:G2G3}
\end{figure}

For $\kappa_l=0.152$ we have computed a 60-noise estimate for each of our
2000 configurations, and for $\kappa_l=0.154$ we have computed a 200-noise
estimate for 250 configurations.
The vector meson masses for these $\kappa_l$ values are 912(8) MeV and
1066(4) MeV respectively (see Table VI of Ref.~\cite{Gockeler}) which
surround $m_\phi=1019$ MeV so that our data will allow interpolation
to a strange quark loop.
From the lattice simulations, three ratios are constructed,
\begin{equation}\label{RX}
R_X(t,t^\prime,\vec{q}) =
     \frac{G_X^{(3)}(t,t^\prime,\vec{q})G^{(2)}(t^\prime,\vec0)}
          {G^{(2)}(t,\vec0)G^{(2)}(t^\prime,\vec{q})},
\end{equation}
where $R_S$, $R_M$ and $R_E$ correspond to the scalar, magnetic and electric
cases respectively, $t$ is the sink timestep and $t^\prime$ the current
insertion timestep.  The two-point and three-point correlators are shown
diagramatically in Fig.~\ref{fig:G2G3}.

Strangeness matrix elements are extracted from the ratios of 
Eq.~(\ref{RX}). Denoting the matrix elements by $M$ with an
obvious subscript, these are related to form factors by
\begin{equation}
M_{\{S,M,E\}}(t,\vec{q}) =
\left\{G_S^{(s)},\frac{\epsilon_{ijk}q_kG_M^{(s)}}{E_q+m},G_E^{(s)}\right\}.
\end{equation}
In the magnetic case, $i$, $j$ and $k$ run over spatial directions
and the corresponding indices on $M_M$ are suppressed for notational
simplicity.

There are various ways in which the matrix element can be extracted from the
ratio. For example, one can sum the contributions for the strange quark 
inserted at different times $t'$. One way\cite{Viehoff} to do this is
\begin{equation}
\sum_{t^\prime=1}^{t}R_X(t,t^\prime,\vec{q}) \to \text{constant}
 + tM_X(t,\vec{q}). \label{cumu1}
\end{equation}
A disadvantage of this kind of method is that the matrix element does
not emerge directly. A fit to the time dependence, which in practice
may be linear only over a limited range, is required to determine 
$M_X$. For this reason we prefer a differential method\cite{Wilcox}
\begin{equation}
\sum_{t^\prime=1}^{t+1}\left[R_X(t,t^\prime,\vec{q})-R_X(t-1,t^\prime,\vec{q})
\right] \to M_X(t,\vec{q}) \label{differential}
\end{equation}
which gives $M_X$ directly. For completeness we also consider the relation
\begin{equation}
\sum_{t^\prime=1}^{t_{fixed}}R_X(t,t^\prime,\vec{q}) \to \text{constant}
 + tM_X(t,\vec{q}),~~~\text{with~}t_{fixed}>t \label{cumu2}
\end{equation}
used in Ref.~\cite{GMKentucky}.

Finally one has to relate the lattice matrix element to the continuum one.
The physical scalar density requires wavefunction
renormalization and we use the tadpole-improved factor,\cite{Ztadpole}
\begin{equation}
\left<N\left|\bar{s}s\right|N\right> = \left(1-\frac{3\kappa_v}
                                       {4\kappa_c}\right)G_S^{(s)},
\end{equation}
with $\kappa_c=0.157096(28)^{+33}_{-9}$\cite{QCDPAX}.
The conserved vector current was used for $G_M^{(s)}$ and $G_E^{(s)}$,
and its normalization is such that no wavefunction renormalization
factor is required.

\begin{figure}[thb]
\includegraphics[width=150mm]{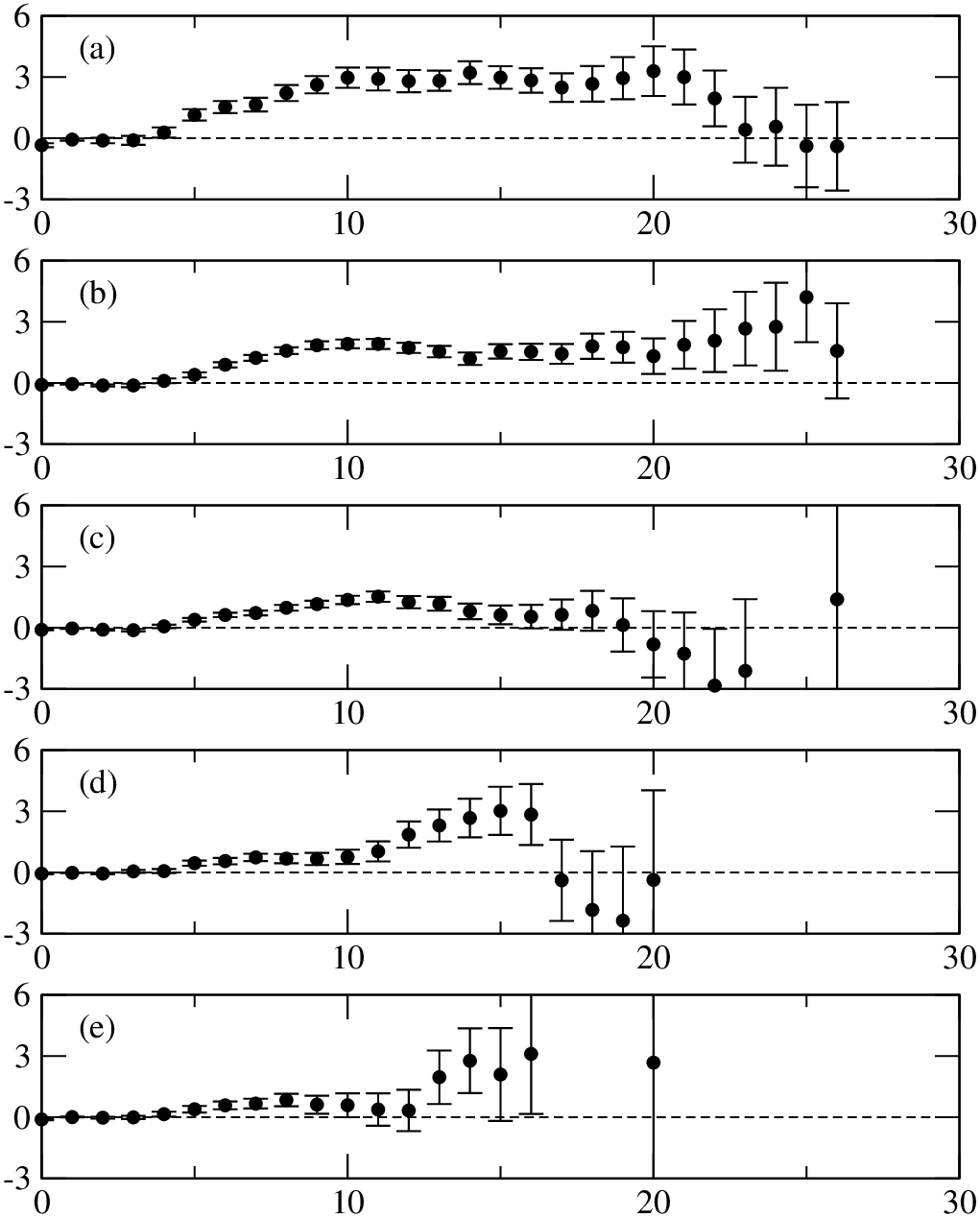}
\caption{Lattice data for the strangeness scalar density as obtained from
         Eq.~(\ref{differential}) with $\kappa_v=0.154$ and $\kappa_l=0.152$.
         Panels (a) through (e) correspond to momenta
         $a^2\vec{q}^{\,2}=n(\pi/10)^2$ with $n=0$ through 4 respectively.
         Uncertainties are calculated from 3000 bootstrap ensembles.
         }\label{fig:scalar}
\end{figure}

\begin{figure}[thb]
\includegraphics[width=150mm]{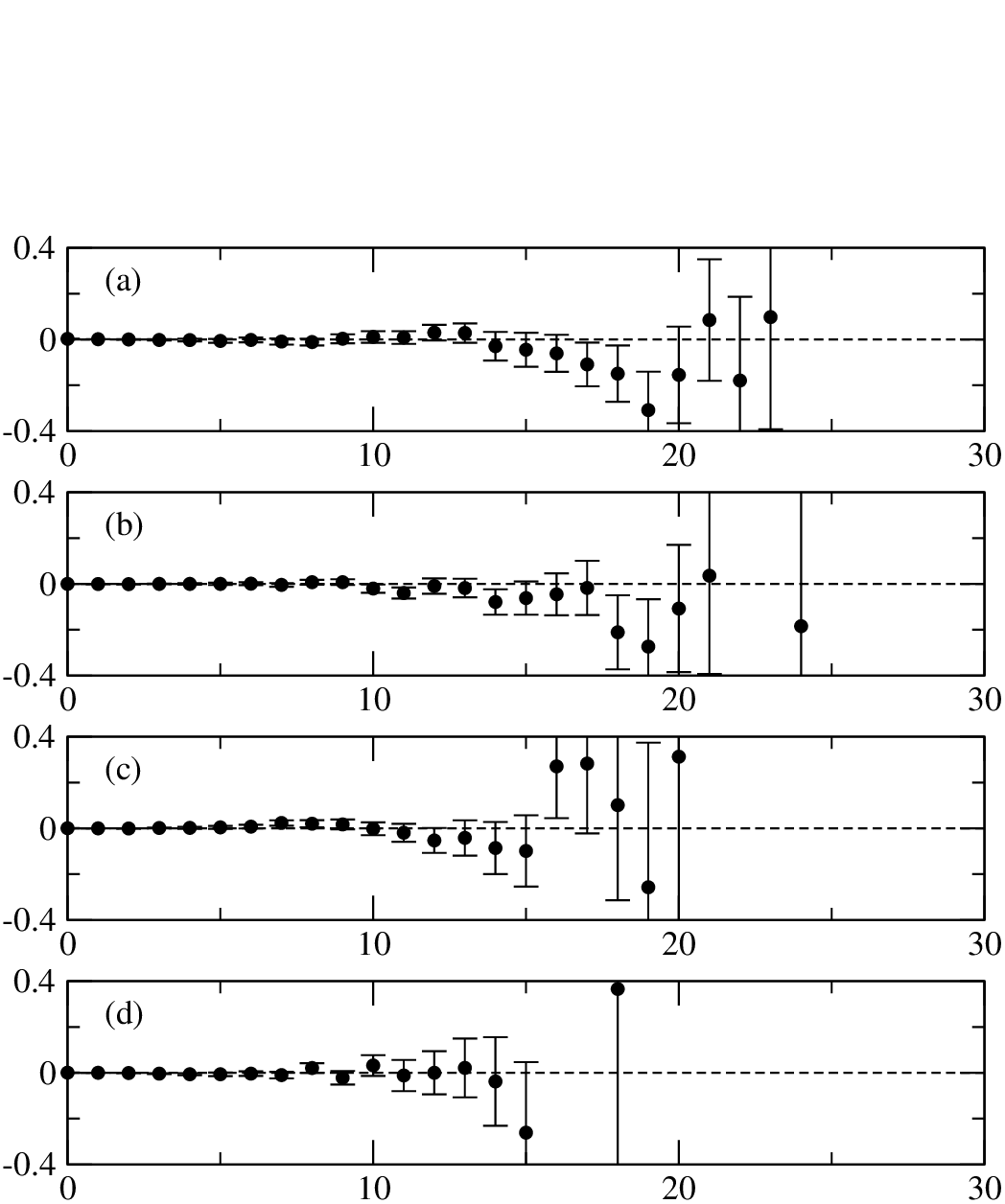}
\caption{Lattice data for the strangeness magnetic form factor as obtained from
         Eq.~(\ref{differential}) with $\kappa_v=0.154$ and $\kappa_l=0.152$.
         Panels (a) through (d) correspond to momenta
         $a^2\vec{q}^{\,2}=n(\pi/10)^2$ with $n=1$ through 4 respectively.
         Uncertainties are calculated from 3000 bootstrap ensembles.
         }\label{fig:magnetic}
\end{figure}

\begin{figure}[thb]
\includegraphics[width=150mm]{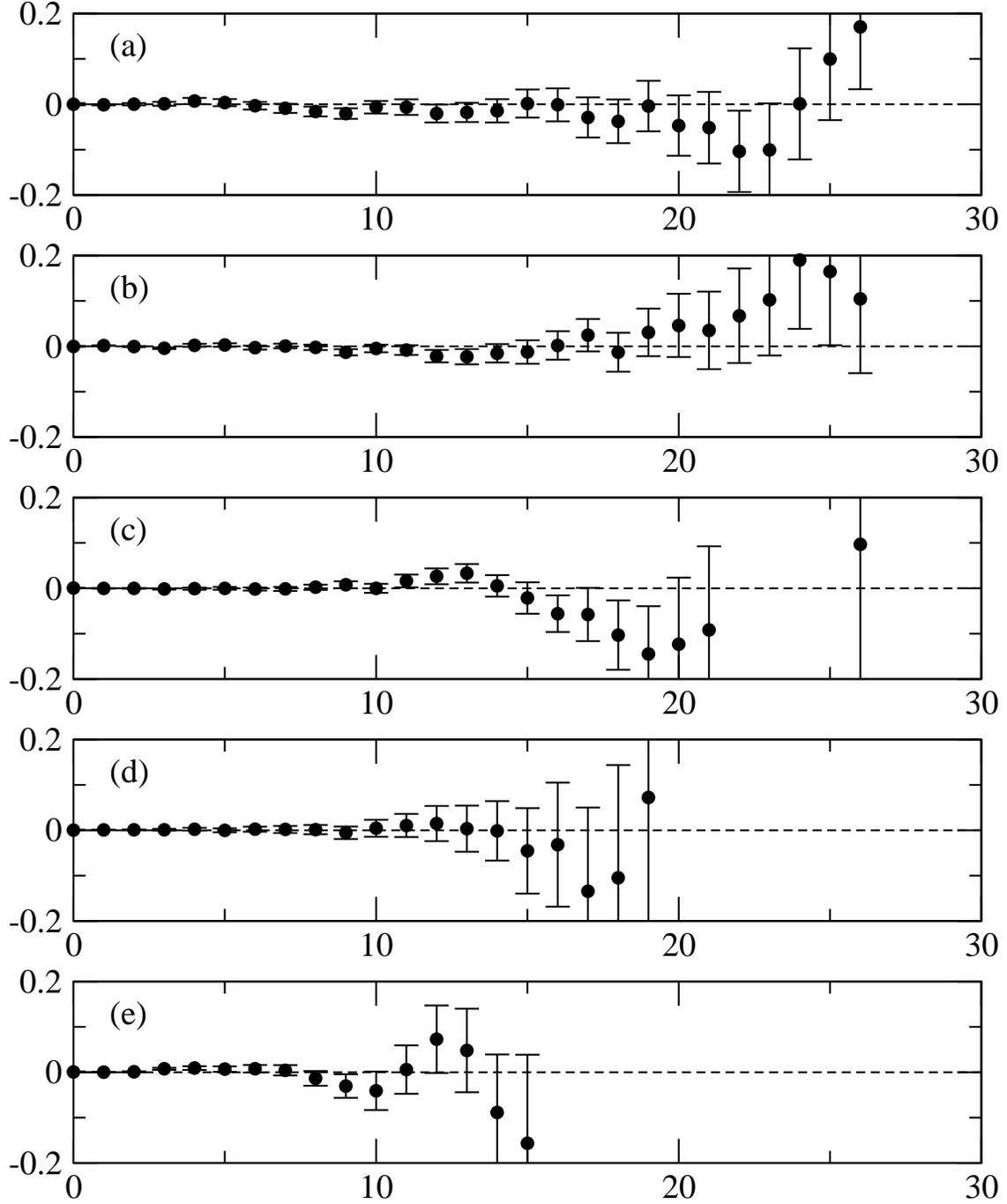}
\caption{Lattice data for the strangeness electric form factor as obtained from
         Eq.~(\ref{differential}) with $\kappa_v=0.154$ and $\kappa_l=0.152$.
         Panels (a) through (e) correspond to momenta
         $a^2\vec{q}^{\,2}=n(\pi/10)^2$ with $n=0$ through 4 respectively.
         Uncertainties are calculated from 3000 bootstrap ensembles.
         }\label{fig:electric}
\end{figure}

Fig.~\ref{fig:scalar} shows our lattice data for the scalar density 
versus timestep,
with $\kappa_v=0.154$ and $\kappa_l=0.152$, analyzed using
Eq.~(\ref{differential}).  In this case there is a very clear
signal and, for each value of the momentum transfer, 
the plateau begins about ten timesteps from the source, although
uncertainties grow with $\vec{q}^{\,2}$.
Figs.~\ref{fig:magnetic} and \ref{fig:electric} show the magnetic and
electric data from Eq.~(\ref{differential}) with the same
$\kappa_v, \kappa_l$ values.  In contrast to the scalar density,
there is no apparent nonzero signal.
However, using the scalar density results, which suggest that the 
plateau region begins about ten timesteps from the source, as a guide,
one concludes
that the form factors are consistent with zero within uncertainties less than
0.1 for all $\vec{q}^{\,2}$ values studied.  We have verified that
Eqs.~(\ref{cumu1}) and (\ref{cumu2}) produce compatible results for all
three matrix elements.

The results of fitting each of our lattice measurements to
Eq.~(\ref{differential}) over four consecutive timesteps,
beginning ten timesteps from the source in every
case, are tabulated in Table~\ref{tab:raw} with statistical uncertainties 
obtained from a bootstrap analysis employing 3000 bootstrap ensembles.
If the uncertainties simply scaled with the square root of
the number of configurations then the ratio of uncertainties between
$\kappa_v=0.154$ and 0.152 should be near 2.8, but the increased number of
noises per configuration for $\kappa_v=0.154$ could reduce this ratio.
According to Table~\ref{tab:raw}, only $G_M^{(s)}$ shows a noticeable
dependence on the number of noises.

These results for $G_M^{(s)}$ can be compared to the findings of
Ref.~\cite{GMKentucky}, since
those authors also work with the Wilson action with the same $\beta$ and
$\kappa$ values, although their lattice volume is smaller.
{}From 100 configurations with 300 complex $Z_2$ noises analyzed using
the method of Eq.~(\ref{cumu2}) only, those authors
interpreted their results to imply a nonzero value for $G_M^{(s)}$.  Our
studies (see Ref.~\cite{emi} for a specific discussion) suggest that a clearer
picture is attained with a larger sample of gauge configurations.  According to
Table~\ref{tab:raw}, even the small statistical uncertainties of the
present work do not permit a definitive nonzero determination of $G_M^{(s)}$.
The same is true for $G_E^{(s)}$.

\section{Chiral extrapolations}

\begin{figure}[thb]
\includegraphics[width=150mm]{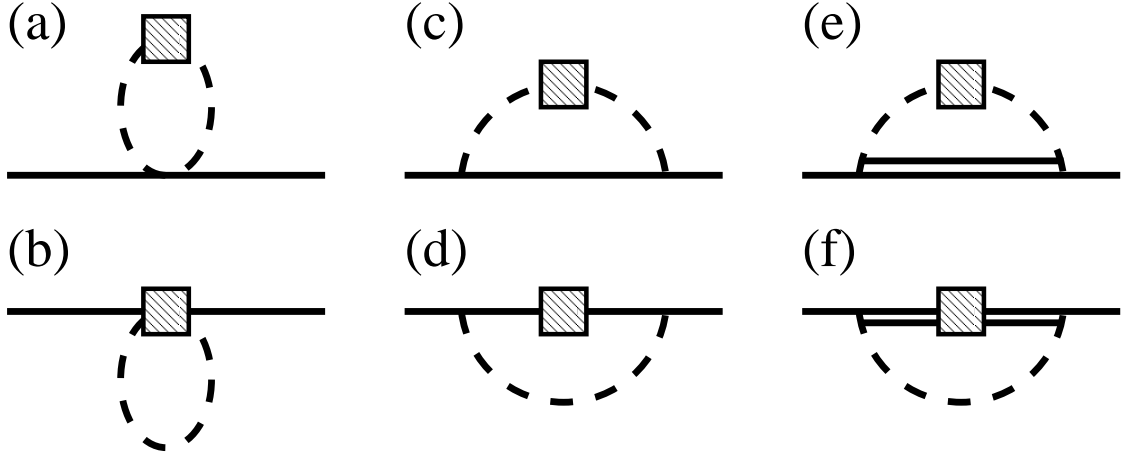}
\caption{Leading loop diagrams for the strangeness matrix elements from
         quenched chiral perturbation theory.
         Dashed, solid and double lines denote octet mesons, octet baryons
         and decuplet baryons respectively.  A shaded box denotes a current
         insertion.}\label{fig:loops}
\end{figure}

Consider quenched SU(3) ChPT with explicit fields for the
pseudoscalar meson octet ($M$), spin-1/2
baryon octet ($B$), spin-3/2 baryon decuplet ($T$) and external electromagnetic
and scalar fields.  The ChPT Lagrangian is
\begin{equation}
{\cal L} = {\cal L}_{M}^{(2)} + {\cal L}_{MB}^{(0)}
         + {\cal L}_{MB}^{(1)} + {\cal L}_{MB}^{(2)}
         + {\cal L}_{MB}^{(3)} + {\cal L}_{MT}^{(1)}
         + {\cal L}_{MBT}^{(1)} + \ldots,
\end{equation}
where a superscript ``$(n)$'' denotes an $n$th order contribution from the
expansion in the smaller scales --- momentum transfer, meson masses and the
$T$-$B$ mass splitting $\Delta$ --- relative to the larger
scales $\Lambda_\chi\approx4\pi F_\pi$ and baryon masses.
The leading loop diagrams for our three strangeness form factors
begin at third order and are displayed in Fig.~\ref{fig:loops}.
Each diagram receives contributions from various quark flows
which have been calculated using the approach of Labrenz and
Sharpe\cite{LabSha}.
Besides these loop contributions, there are also contact terms in the
Lagrangian which contribute low energy constants (LEC's) to the strangeness
matrix elements.
Here are the explicit formulae:
\begin{eqnarray}
\left<N\left|\bar{s}s\right|N\right> &=& C_1\mu + C_2^r(\lambda)\mu\Delta 
              - \frac{\mu\pi C_B}{4(4\pi F_\pi)^2}\left[4m_K+
            \int_0^1\text{d}x\frac{(2m_K^2-q^2)}{\sqrt{m_K^2-x(1-x)q^2}}\right]
           \nonumber \\
          &&  - \frac{\mu\pi\gamma^2}{2(4\pi F_\pi)^2}\left[4m_{\bar{s}s}+
            \int_0^1\text{d}x\frac{(2m_{\bar{s}s}^2-q^2)}
            {\sqrt{m_{\bar{s}s}^2-x(1-x)q^2}}\right]
           \nonumber \\
          && - \frac{\mu\Delta C_T}{2(4\pi F_\pi)^2}\left[-\ln\left(\frac{m_K^2}
                {\lambda^2}\right) - \int_0^1\text{d}x\,\ln\left(1-x(1-x)
                \frac{q^2}{m_K^2}\right) \right.\nonumber \\
          &&\left.  +\frac{2}{\Delta}\int_0^1\text{d}x\left(\frac{\Delta^2
                -m_K^2+(4/3)x(1-x)q^2}{\Delta^2-m_K^2+x(1-x)q^2}\right)
                A(x)\right], \label{SChPT} \\
G_M^{(s)}(q^2) &=& C_3 + C_4^r(\lambda)\Delta 
             + \frac{2\pi m_NC_B}{(4\pi F_\pi)^2}\int_0^1
             {\text{d}x}\,\sqrt{m_K^2-x(1-x)q^2} \nonumber \\
          && + \frac{m_N\Delta C_T}{3(4\pi F_\pi)^2}\left[\ln\left(\frac{m_K^2}{
             \lambda^2}\right) - \frac{11}{3} - \frac{2}{\Delta}\int_0^1
             \text{d}x\,A(x)
             - \int_0^1\text{d}x\frac{2m_K^2-q^2/2}{m_K^2-x(1-x)q^2} \right],
          \nonumber \\ &&   \label{MChPT} \\
G_E^{(s)}(q^2) &=& C_5^r(\lambda)q^2 + C_6\frac{q^2}{m_N} \nonumber \\
          && + \frac{2q^2C_B}{3(4\pi F_\pi)^2}\left[\frac{5}{8}
             \ln\left(\frac{m_K^2}{\lambda^2}\right)+\frac{17}{24}-
             \left(\frac{m_K^2}{q^2}-\frac{5}{8}\right)\int_0^1\text{d}x
             \,\ln\left(1-x(1-x)\frac{q^2}{m_K^2}\right)\right] \nonumber \\
          && +\frac{q^2}{4(4\pi F_\pi)^2}\left[\ln\left(\frac{m_K^2}
             {\lambda^2}\right) + \frac{1}{3} + \left(1-
             \frac{4m_K^2}{q^2}\right)\int_0^1\text{d}x\,\ln\left(1-x(1-x)
             \frac{q^2}{m_K^2}\right)\right] \nonumber \\
          && + \frac{C_Tq^2}{(4\pi F_\pi)^2}\left[\frac{5}{36}\ln\left(
             \frac{m_K^2}{\lambda^2}\right)+\frac{m_K^2}{9q^2}-\frac{2\Delta^2}
             {q^2}-\frac{7}{54}+\frac{2\Delta}{q^2}A(0) \right.\nonumber \\
          && -\left(\frac{m_K^4}
             {9q^2}-\frac{2m_K^2\Delta^2}{q^2}-\frac{7m_K^2}{18}
             +\frac{\Delta^2}{2}+\frac{5q^2}{72}\right)
             \int_0^1\frac{\text{d}x}{m_K^2-x(1-x)q^2} \nonumber \\
          && \left. 
             - \frac{2\Delta}{q^2}\int_0^1\text{d}x\left(\frac{\Delta^2-m_K^2
             +(4/3)x(1-x)q^2}
             {\Delta^2-m_K^2+x(1-x)q^2}\right)A(x)\right],
\end{eqnarray}
where $\Delta>0$ and
\begin{eqnarray}
A(x) &=& \left\{\begin{array}{ll}
         \sqrt{z-\Delta^2}\arccos\left(\frac{\Delta}{\sqrt{z}}\right)
         & \text{~for~} \Delta<\sqrt{z}, \\
         -\sqrt{\Delta^2-z}\ln\left(\frac{\Delta}{\sqrt{z}}
         +\sqrt{\frac{\Delta^2}{z}-1}\right)
         & \text{~for~} \Delta>\sqrt{z},
         \end{array}\right.
\end{eqnarray}
with
\begin{equation}
z \equiv m_K^2-x(1-x)q^2.
\end{equation}
Our interest is in spacelike $q^2$, so $z$ is positive definite throughout the
range $0<x<1$.  
The $q^2$ of each lattice data point is obtained from
\begin{equation}
q^2 = (E_n-E_0)^2 - n\left(\frac{\pi}{10a}\right)^2
\end{equation}
where $n=0$, 1, 2, 3 or 4 and the $E_n$ are taken from Table~\ref{tab:2point}.

$C_B$ contains the familiar axial couplings ($D$ and $F$)
and $C_T$ contains the octet-decuplet coupling $\cal C$ 
(defined, for example, in Ref.~\cite{LabSha}):
\begin{eqnarray}
C_B &=& \frac{5}{3}D^2-2DF+3F^2, \\
C_T &=& {\cal C}^2.
\end{eqnarray}
The parameters $C_1$, $C_2$, \ldots $C_6$ are LEC's, some of which depend on
the dimensional regularization scale $\lambda$ such that
the full matrix elements are independent of $\lambda$.
$\gamma$ is the ChPT parameter for the quenched $\eta^\prime$\cite{LabSha} and
$m_{\bar{s}s}$ is the mass of a doubly-strange pseudoscalar meson.
The normalization convention corresponds to $F_\pi\approx93$ MeV and
$\mu$ is the ChPT parameter defined by
\begin{equation}
m_K^2 = \mu(\hat{m}+m_s)
\end{equation}
with $\hat{m}\equiv m_u=m_d$.
In order to verify various aspects of these ChPT expressions for
$\left<N\left|\bar{s}s\right|N\right>(q^2)$,
$G_M^{(s)}(q^2)$ and $G_E^{(s)}(q^2)$, comparisons were made to
the collection of papers in Ref.~\cite{ChPTbits}.

Notice that the three strangeness matrix elements contain a total of
six parameters --- $\mu C_1+\mu\Delta C_2^r(\mu)$,
$C_3+\Delta C_4^r(\mu)$, $C_5+C_6^r(\mu)/m_N$, $C_B$, $C_T$ and $\gamma^2$
--- and the dependences on each of these parameters are linear.
Because $D$, $F$, $\cal C$ and $\gamma$ are real parameters it follows
that $C_B$, $C_T$ and $\gamma^2$ must be positive definite, and
Eq.~(\ref{SChPT}) therefore requires that
$\left<N\left|\bar{s}s\right|N\right>(q^2)$ 
decreases as $\hat{m}$, $m_s$ or $-q^2$ is increased.
This is consistent with the lattice QCD data of Table~\ref{tab:raw}.

It should be noted that the range of $\vec{q}$ used in our lattice
simulations extends far beyond the range of applicability of ChPT,
and there is therefore no reason to expect that the form of ChPT
will look anything like the lattice data for these larger momentum
values.  As would be hoped, use of only the lattice data at smaller
momentum values leads to a good ChPT fit.
As it happens, the ChPT expressions fit all three matrix
elements surprisingly well over the entire momentum range studied.
Although this is surely accidental, it means that the ChPT expressions
can be used as a convenient method of smoothly interpolating the
momentum dependences of these matrix elements.

To determine numerical values for the six parameters appearing in the
ChPT expressions, we perform
a least squares fit to the data of Table~\ref{tab:raw}.
In particular, we'll fit the 39 data points having $\kappa_l=0.152$
(data for $G_E^{(s)}(0)$ are omitted since gauge invariance requires a
zero result) and verify that predictions for $\kappa_l=0.154$ are
consistent with our lattice simulations.
We will also perform an independent fit using only 12 of the
39 data points: those having $a^2\vec{q}^2=0$ or $a^2\vec{q}^2=(\pi/10)^2$.
These smallest momenta are the ones most appropriate to ChPT and, as
will be demonstrated, the final predictions for strangeness matrix elements
are rather insensitive to whether or not the higher momentum data
are used as input for the ChPT fit.
The statistical uncertainties of the fit parameters are determined from 
a bootstrap analysis.

In addition to the statistical error there is a systematic uncertainty
due to the choice of chiral model.
The dynamics of the ChPT expressions reside in the loop diagrams, and they
contain the quenched $\eta^\prime$ parameter $\gamma^2$ as well as the
non-$\eta^\prime$ parameters $C_B$ and $C_T$.
It is possible to obtain a good fit to the $\kappa_l=0.152$ data in the
extreme limit of no $\eta^\prime$ ($\gamma^2=0$) or in the opposite extreme
of ``maximal $\eta^\prime$'' where $C_B=C_T=0$.  (In the maximal $\eta^\prime$
case, we also choose $C_3+C_4^r(\lambda)\Delta=0$
since it is clear from Eq.~(\ref{MChPT}) that this parameter would simply
be an additive constant for $G_M^{(s)}(q^2)$ and would be consistent with
zero when fitted to our lattice QCD data.)
These separate possibilities indicate that our lattice data are not
precise enough to determine the
fraction of $\eta^\prime$ physics in the strangeness form factors.
One might expect
the physical values for these parameters to lie somewhere between the
two extremes, and we will use this range to define a theoretical
error bar.
The results of our fits to the $\kappa_l=0.152$ data, and the resulting
predictions for $\kappa_l=0.154$, are recorded in Table~\ref{tab:fit}.
The fits are consistent with the direct lattice QCD simulations of
Table~\ref{tab:raw}.
The corresponding ChPT parameter values are listed in Table~\ref{tab:fitpars},
along with the parameter values obtained from fits to the data having
the two smallest momenta: $a^2\vec{q}^2=0$ and $a^2\vec{q}^2=(\pi/10)^2$.
In the unquenched theory $\gamma^2$ does not appear and standard phenomenology
leads to $C_B\sim0.9$ and $1.4\alt C_T\alt2$.
Not surprisingly, the quenched parameter values in
Table~\ref{tab:fitpars} are different but are still $O(1)$.

\begin{figure}[thb]
\includegraphics[width=138mm]{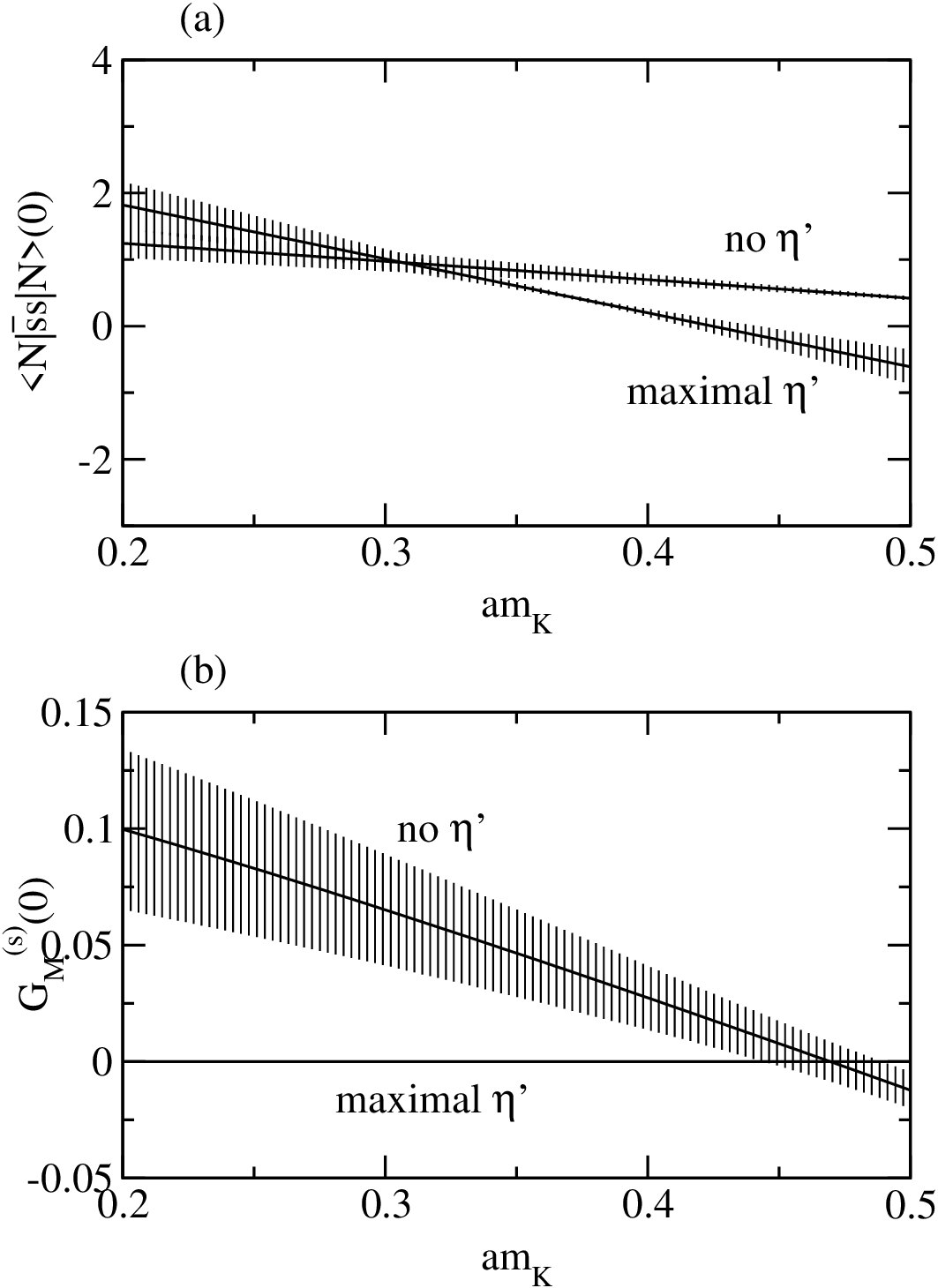}
\caption{Strangeness matrix elements at $-q^2=0$ as functions of $m_K$.
         The two solid curves represent the extreme cases of maximizing or
         minimizing the quenched $\eta^\prime$ contributions in ChPT loop
         diagrams relative to non-$\eta^\prime$ physics.  ChPT parameters
         are obtained from a fit to 39
         lattice QCD data points as discussed in the text, and the thickness
         of a hatched band denotes statistical uncertainties from 3000 bootstrap
         ensembles.}\label{fig:vsmK39}
\end{figure}

\begin{figure}[thb]
\includegraphics[width=138mm]{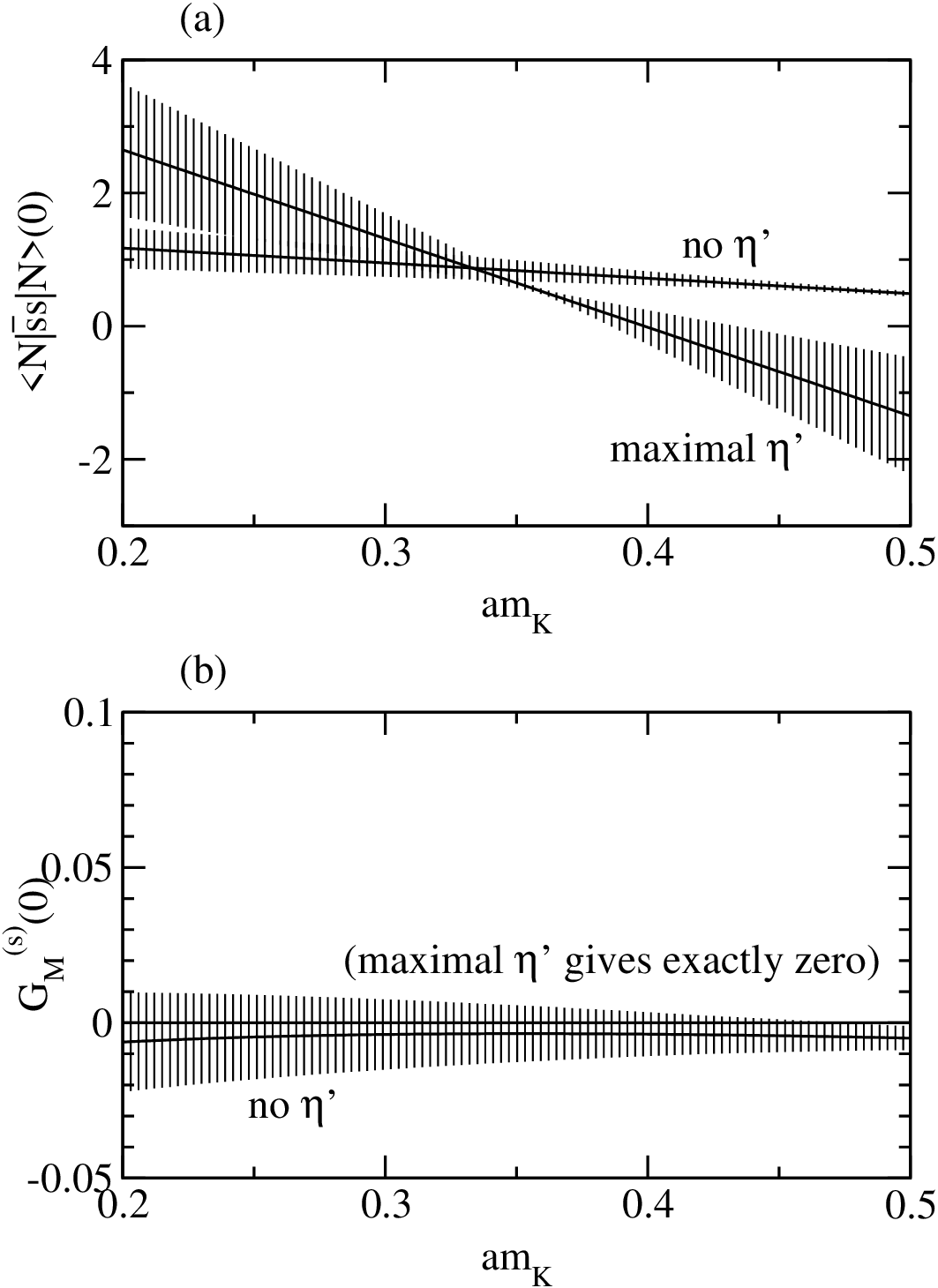}
\caption{Strangeness matrix elements at $-q^2=0$ as functions of $m_K$.
         The two solid curves represent the extreme cases of maximizing or
         minimizing the quenched $\eta^\prime$ contributions in ChPT loop
         diagrams relative to non-$\eta^\prime$ physics.  ChPT parameters
         are obtained from a fit to 12
         small-momentum lattice QCD data points ($a^2\vec{q}^2=0$ and
         $a^2\vec{q}^2=(\pi/10)^2$) as discussed in the text, and the thickness
         of a hatched band denotes statistical uncertainties from 3000 bootstrap
         ensembles.}\label{fig:vsmK12}
\end{figure}

For physical meson masses,
\begin{equation}
\frac{\hat{m}}{m_s} = \frac{m_\pi^2}{2m_K^2-m_\pi^2} = \frac{1}{25}
\end{equation}
which leads to
\begin{equation}
m_{\bar{s}s} = m_K\sqrt{\frac{2}{1+\hat{m}/m_s}} = 1.39m_K.
\end{equation}
At $-q^2=0$, $G_E^{(s)}$ vanishes identically.  Figs.~\ref{fig:vsmK39} and
\ref{fig:vsmK12} show the
other two strangeness matrix elements as functions of the kaon mass.
Fixing $m_K$ to its physical value leads to the momentum dependent
strangeness matrix elements of Figs.~\ref{fig:vsqq39} and \ref{fig:vsqq12},
which are our final
results.  Comparison to experiment, along with disclaimers about such a
comparison, are contained in Section~\ref{sec:discussion}.

To conclude this section we return to the suggestion from Ref.~\cite{ChenSav}
of using SU(2) ChPT instead of SU(3). This is an appealing idea because
SU(2) ChPT typically converges more rapidly.
In effect, the kaon loop diagrams of Fig.~\ref{fig:loops} get replaced
by SU(2) LEC's.  Although SU(3) ChPT uses a common set of parameters
($C_B$, $C_T$ and $\gamma^2$) for the kaon loop effects in all three
strangeness matrix elements, SU(2) ChPT has separate LEC's
for each matrix element.  Since the raw lattice QCD data of Table~\ref{tab:raw}
only reveal a nonzero signal for the strangeness scalar density, it is
difficult to discuss SU(2) ChPT extrapolations of the strangeness
electromagnetic form factors in any detail.  Perhaps future lattice QCD data
for these form factors will be precise enough to benefit from SU(2) ChPT.

\section{Discussion}\label{sec:discussion}

\begin{figure}[thb]
\includegraphics[width=138mm]{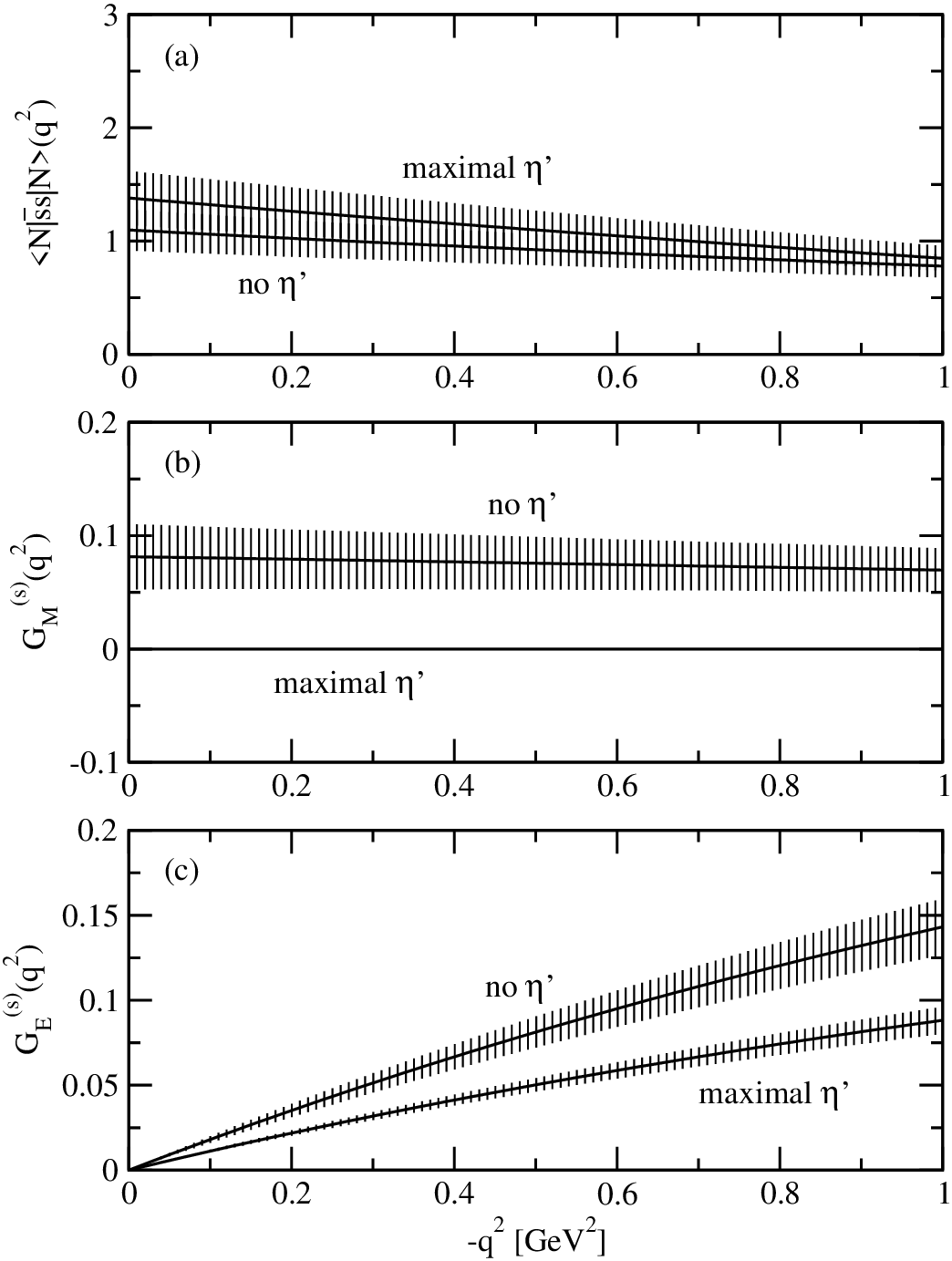}
\caption{Strangeness matrix elements as functions of $-q^2$.
         The two solid curves represent the extreme cases of maximizing or
         minimizing the quenched $\eta^\prime$ contributions in ChPT loop
         diagrams relative to non-$\eta^\prime$ physics.  ChPT parameters
         are obtained
         from a fit to 39 lattice QCD data points as discussed in the text, and
         the thickness of a hatched band denotes statistical uncertainties from 3000
         bootstrap ensembles.}\label{fig:vsqq39}
\end{figure}

\begin{figure}[thb]
\includegraphics[width=138mm]{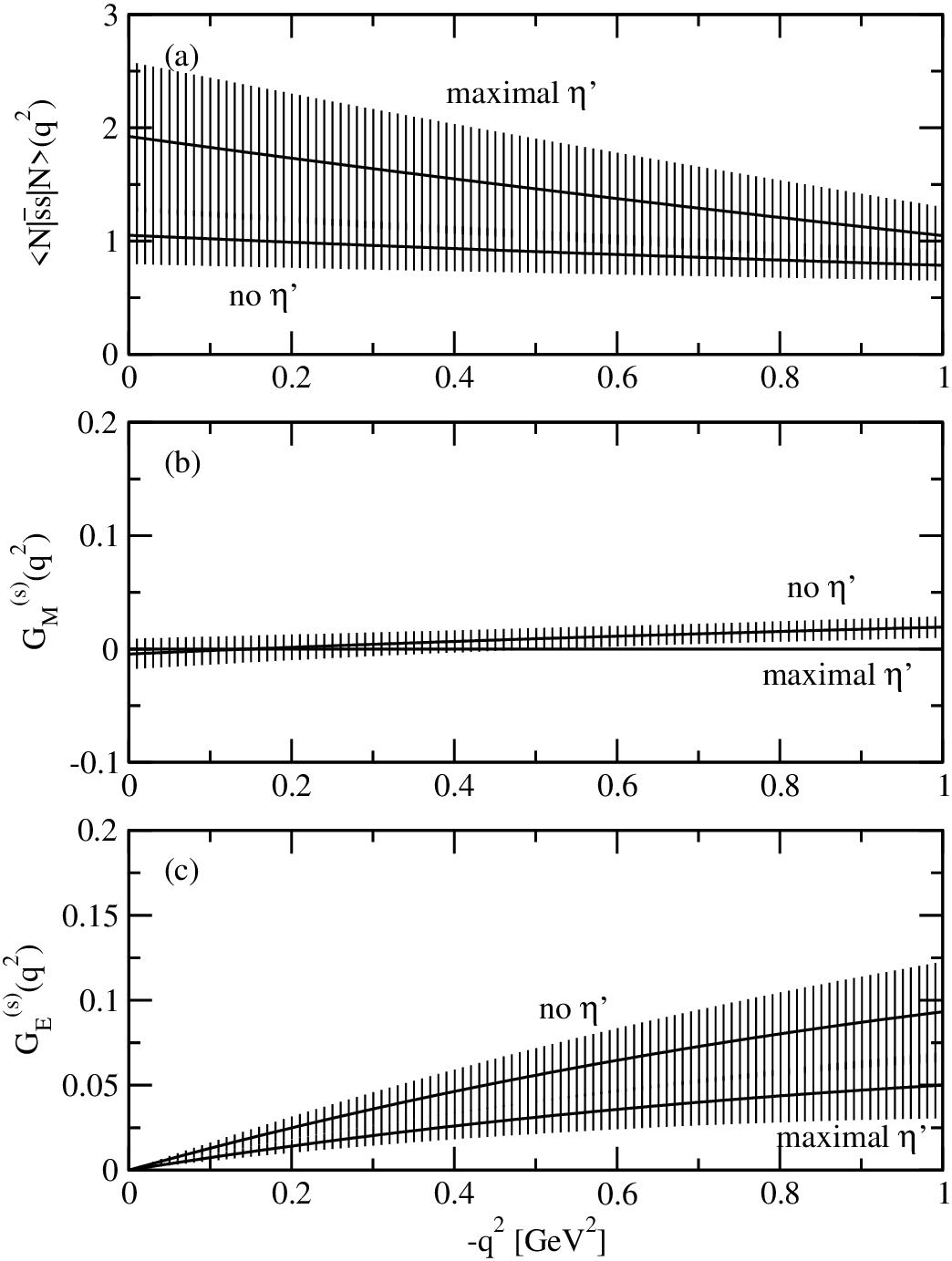}
\caption{Strangeness matrix elements as functions of $-q^2$.
         The two solid curves represent the extreme cases of maximizing or
         minimizing the quenched $\eta^\prime$ contributions in ChPT loop
         diagrams relative to non-$\eta^\prime$ physics.  ChPT parameters
         are obtained from a fit to 12 small-momentum lattice QCD data points
         ($a^2\vec{q}^2=0$ and $a^2\vec{q}^2=(\pi/10)^2$) as discussed in
         the text, and the thickness of a hatched band denotes statistical uncertainties
         from 3000 bootstrap ensembles.}\label{fig:vsqq12}
\end{figure}

The results of this work (Figs.~\ref{fig:vsqq39} and \ref{fig:vsqq12})
compare favourably to the available experimental data:
\begin{eqnarray}
G_M^{(s)}(q_1^2) &=& \left\{\begin{array}{ll}
 0.14 \pm 0.29 \pm 0.31, & \text{Ref.~\cite{SAMPLE}}, \\
 0.05 \pm 0.06,          & \text{This work},
   \end{array}\right. \\
G_E^{(s)}(q_2^2) + 0.39G_M^{(s)}(q_2^2) &=& \left\{\begin{array}{ll}
 0.025 \pm 0.020 \pm 0.014, & \text{Ref.~\cite{HAPPEX}}, \\
 0.07 \pm 0.05,           & \text{This work},
   \end{array}\right.
\end{eqnarray}
where $-q_1^2 = 0.1 \text{~GeV}^2$ and $-q_2^2 = 0.477 \text{~GeV}^2$.
Here, the uncertainties (incorporating both statistical and theoretical
modeling errors) 
in our results have been estimated by the requirement
that all curves from
Figs.~\ref{fig:vsqq39} and \ref{fig:vsqq12}, representing fits to all momenta,
fits to only small
momenta, ``maximal $\eta^\prime$'' fits and ``no $\eta^\prime$'' fits are
within one standard deviation of the quoted central value.
The lack of a fundamental scalar probe makes the strangeness scalar density
harder to extract from experiment, but  Figs.~\ref{fig:vsqq39} and
\ref{fig:vsqq12} can be compared to other quenched lattice QCD simulations.
The renormalization group
invariant quantity representing the fractional strange quark contribution to
the nucleon mass is:
\begin{equation}
\frac{m_s\left<N\left|\bar{s}s\right|N\right>(0)}{m_N} = \left\{
\begin{array}{lll} 0.302(48) & \text{~at~}\beta=5.7 & \text{Ref.~\cite{FKOU}},\\
                   0.195(9) & \text{~at~}\beta=6.0 & \text{Ref.~\cite{DLL}}, \\
                   0.21(11) & \text{~at~}\beta=6.0 & \text{This work}.
\end{array}\right.
\end{equation}
If the curves of Fig.~\ref{fig:vsqq12} are not included in the predictions
for these strangeness matrix elements and if the statistical errors of
Fig.~\ref{fig:vsqq39} are ignored relative to the theoretical errors
(reflecting the difference between ``maximal $\eta^\prime$'' and
``no $\eta^\prime$'' fits), then one arrives at the earlier
results reported in Ref.~\cite{lat02}:
$G_M^{(s)}(q_1^2) = 0.03 \pm 0.03$,
$G_E^{(s)}(q_2^2) + 0.39G_M^{(s)}(q_2^2) = 0.027 \pm 0.016$ and
$(m_s/m_N)\left<N\left|\bar{s}s\right|N\right>(0) = 0.15(2)$.

There are a number of ways that future theoretical studies could improve
upon the results obtained in this work.  From the outset we have
restricted ourselves to the quenched approximation, and this introduces a
systematic error that is perhaps 10-20\%\cite{quenching}.
It is also not obvious that higher orders in the ChPT expansion are small
for the case at hand, i.e.
SU(3) ChPT for baryons with quark masses in the strange
region.  It would be interesting to see the results of partially quenched
simulations and lighter valence quarks for these strangeness matrix elements.
Refinements of the disconnected loop techniques could also be advantageous,
such as perturbative subtraction beyond $O(\kappa^4,\kappa^5)$
and heatbath noise methods\cite{hotnoise}.  Finally, we recall that the
so-called strangeness electric and magnetic form factors would not be
exactly zero even in a world without any strange quark, due to isospin
violation\cite{isospin,LewMob}.  Based on Ref.~\cite{LewMob}, the isospin
violation effects are not so different in magnitude from the tiny
strange quark effects discussed in the present work.

Although there are certainly further steps that can be taken toward a
more detailed understanding of these strangeness matrix elements,
the present study has established that $G_E^{(s)}(q^2)$ and $G_M^{(s)}(q^2)$
are small over the range of momenta and quark masses used in these lattice
QCD simulations, and that they remain small when extrapolated with quenched
SU(3) ChPT in combination with lattice QCD data for
$\left<N\left|\bar{s}s\right|N\right>(q^2)$.

\section*{Acknowledgments}

This work was supported in part by the National Science Foundation under
grant 0070836, the Baylor Sabbatical Program, and the Natural Sciences and
Engineering Research
Council of Canada.  Some of the computing was done on hardware funded by the
Canada Foundation for Innovation with contributions from Compaq Canada,
Avnet Enterprise Solutions and the Government of Saskatchewan.

\begin{table}[thb]
\caption{Dimensionless energies of the nucleon, $aE_n$, with momentum squared
$a^2\vec{q}^{\,2} = n(\pi/10)^2$.  All fits begin 16 timesteps from
the source, 2000 configurations are used, and statistical uncertainties are from 
a bootstrap analysis with 3000 bootstrap ensembles.}\label{tab:2point}
\begin{tabular}{cr@{.}lr@{.}lr@{.}l}
\hline\hline
$n$ & \multicolumn{2}{c}{$\kappa=0.152$} &
 \multicolumn{2}{c}{$\kappa=0.153$} & \multicolumn{2}{c}{$\kappa=0.154$} \\
\hline
\hline
0 & 0&869(2)  & 0&799(2)  & 0&728(3)  \\
1 & 0&927(3)  & 0&862(3)  & 0&795(4)  \\
2 & 0&986(4)  & 0&924(5)  & 0&865(7)  \\
3 & 1&034(7)  & 0&977(10) & 0&922(15) \\
4 & 1&070(12) & 1&013(18) & 0&945(30) \\
\hline
\hline
\end{tabular}
\vspace{10cm}
\end{table}

\newpage

\begin{table}[thb]
\caption{Fits to the matrix elements of Eq.~(\protect\ref{differential})
         beginning 10 timesteps from the source.  The momentum squared is
         $a^2\vec{q}^{\,2} = n(\pi/10)^2$.  Statistical uncertainties are from 
	 a bootstrap analysis with 3000 bootstrap ensembles.}\label{tab:raw}
\begin{tabular}{ccr@{.}lr@{.}lr@{.}lr@{.}lr@{.}lr@{.}l}
\hline\hline
$\kappa_v$ & $n$ & \multicolumn{6}{c}{$\kappa_l = 0.152$} &
                   \multicolumn{6}{c}{$\kappa_l = 0.154$} \\
\cline{3-14}
 & &
 \multicolumn{2}{c}{$G_S^{(s)}$} & \multicolumn{2}{c}{$G_M^{(s)}$} &
 \multicolumn{2}{c}{$G_E^{(s)}$} & \multicolumn{2}{c}{$G_S^{(s)}$} &
 \multicolumn{2}{c}{$G_M^{(s)}$} & \multicolumn{2}{c}{$G_E^{(s)}$} \\
\hline
\hline
0.152 & 0 & 2&6(4) & \multicolumn{2}{c}{---} & -0&009(13) &
            3&7(13) & \multicolumn{2}{c}{---} & 0&003(5) \\
      & 1 & 1&7(2) &               0&007(16) & -0&008(8)  &
            2&1(6) &              -0&007(15) & -0&027(33)  \\
      & 2 & 1&2(2) &              -0&018(14) &  0&012(10) &
            1&1(6) &               0&008(13) &  0&014(23) \\
      & 3 & 1&1(5) &              -0&014(23) &  0&008(17) &
            1&2(9) &               0&047(41) &  0&017(61) \\
      & 4 & 0&7(6) &               0&004(31) &  0&026(40) &
            3&3(18) &               0&033(59) & -0&046(71) \\
\hline
0.153 & 0 & 2&7(5) & \multicolumn{2}{c}{---} & -0&010(15) &
            4&0(14) & \multicolumn{2}{c}{---} & 0&002(7) \\
      & 1 & 1&8(3) &               0&012(22) & -0&011(10) &
            2&2(7) &              -0&010(17) & -0&034(44) \\
      & 2 & 1&3(2) &              -0&021(20) &  0&015(14) &
            1&2(7) &               0&014(16) &  0&021(32) \\
      & 3 & 1&2(6) &              -0&018(32) &  0&008(22) &
            1&3(11) &               0&071(56) &  0&024(89) \\
      & 4 & 0&7(8) &               0&005(48) &  0&029(56) &
            3&8(22) &               0&049(80) & -0&066(112) \\
\hline
0.154 & 0 & 2&9(5) & \multicolumn{2}{c}{---} & -0&013(19) &
            4&2(15) & \multicolumn{2}{c}{---} & 0&002(9) \\
      & 1 & 1&8(3) &               0&019(33) & -0&014(15) &
            2&3(8) &              -0&016(22) & -0&043(63) \\
      & 2 & 1&3(3) &              -0&022(31) &  0&019(21) &
            1&3(8) &               0&023(23) &  0&032(48) \\
      & 3 & 1&5(9) &              -0&029(53) &  0&008(32) &
            1&4(14) &               0&118(90) &  0&027(149) \\
      & 4 & 0&8(11) &              0&010(82) &  0&021(81) &
            4&5(29) &              0&084(116) & -0&105(191) \\
\hline
\hline
\end{tabular}
\vspace{5cm}
\end{table}

\newpage

\begin{table}[thb]
\caption{Predictions of quenched SU(3) ChPT, after a least squares fit to 39
         lattice data.  The estimated uncertainties include the range between the
         two extreme cases of maximizing or
         minimizing the quenched $\eta^\prime$ contribution in ChPT loop
         diagrams relative to non-$\eta^\prime$ physics,
         as discussed in the text, as well as the statistical
         uncertainties from a bootstrap analysis.}\label{tab:fit}
\begin{tabular}{ccr@{.}lr@{.}lr@{.}lr@{.}lr@{.}lr@{.}l}
\hline\hline
$\kappa_v$ & $n$ & \multicolumn{6}{c}{$\kappa_l = 0.152$} &
                   \multicolumn{6}{c}{$\kappa_l = 0.154$} \\
\cline{3-14}
 & &
 \multicolumn{2}{c}{$G_S^{(s)}$} & \multicolumn{2}{c}{$G_M^{(s)}$} &
 \multicolumn{2}{c}{$G_E^{(s)}$} & \multicolumn{2}{c}{$G_S^{(s)}$} &
 \multicolumn{2}{c}{$G_M^{(s)}$} & \multicolumn{2}{c}{$G_E^{(s)}$} \\
\hline
\hline
0.152 & 0 & 2&1(5) & \multicolumn{2}{c}{---} &  0&0      &
            4&0(18) & \multicolumn{2}{c}{---} & 0&0      \\
      & 1 & 1&7(3) &              -0&006(6)  &  0&002(3) &
            3&4(15) &              0&012(12)  &  0&011(5) \\
      & 2 & 1&3(2) &              -0&006(6)  &  0&001(7) &
            2&8(12) &              0&011(11)  &  0&017(9) \\
      & 3 & 0&9(3) &              -0&007(7)  & -0&003(11) &
            2&2(8) &               0&010(10)  &  0&021(12) \\
      & 4 & 0&6(4) &              -0&008(8)  & -0&008(15) &
            1&9(8) &               0&010(10)  &  0&021(14) \\
\hline
0.153 & 0 & 2&3(4) & \multicolumn{2}{c}{---} &  0&0      &
            4&1(17) & \multicolumn{2}{c}{---} & 0&0      \\
      & 1 & 1&8(2) &               0&008(8) &  0&007(4) &
            3&5(14) &              0&020(20) &  0&017(7) \\
      & 2 & 1&4(2) &               0&007(7) &  0&010(7) &
            3&0(11) &              0&018(18) &  0&028(12) \\
      & 3 & 1&0(4) &               0&006(6) &  0&010(9) &
            2&5(8) &               0&017(17) &  0&036(16) \\
      & 4 & 0&7(5) &               0&005(8) &  0&008(11) &
            2&0(7) &               0&016(16) &  0&041(20) \\
\hline
0.154 & 0 & 2&4(3) & \multicolumn{2}{c}{---} &  0&0     &
            4&2(16) & \multicolumn{2}{c}{---} & 0&0     \\
      & 1 & 2&0(3) &               0&014(14) &  0&012(5) &
            3&6(13) &              0&027(27) & 0&023(9) \\
      & 2 & 1&5(3) &               0&013(13) &  0&019(9) &
            3&1(10) &              0&025(25) & 0&041(15) \\
      & 3 & 1&1(5) &               0&012(12) &  0&024(13) &
            2&8(9) &               0&023(23) & 0&054(21) \\
      & 4 & 0&8(6) &               0&011(11) &  0&025(15) &
            2&2(5) &               0&022(22) & 0&063(25) \\
\hline
\hline
\end{tabular}
\vspace{5cm}
\end{table}

\newpage

\begin{table}[thb]
\caption{The parameter values obtained for the two extreme fits to our
         quenched lattice QCD data at $\kappa_l=0.152$, as discussed in
         the text, using (i) lattice data from all available momenta and
         (ii) lattice data with $a^2\vec{q}^2=0$ and $a^2\vec{q}^2=(\pi/10)^2$
         only.  Statistical uncertainties are from a bootstrap analysis
         with 3000 bootstrap ensembles.}\label{tab:fitpars}
\begin{tabular}{ccccc}
\hline
\hline
           & \multicolumn{2}{c}{(i) fit to all $\vec{q}$}
           & \multicolumn{2}{c}{(ii) fit to small $\vec{q}$} \\
\hline
           & maximal $\eta^\prime$ & no $\eta^\prime$ 
           & maximal $\eta^\prime$ & no $\eta^\prime$ \\
\hline
\hline
$\mu C_1+\mu\Delta C_2^r(1\text{GeV})$ & 3.2(7)  & 1.7(3)  & 5(2)    & 1.6(5) \\
$C_3+\Delta C_4^r(1\text{GeV})$        & ---     & 0.31(7) & ---     & 0.09(4)\\
$C_B$                                  & ---     & 0.11(3) & ---     & 0.12(6)\\
$\gamma^2$                             & 0.45(11)& ---     & 0.7(3)  & --- \\
$C_T$                                  & ---     & 1.0(2)  & ---     & 0.8(3) \\
$[C_5+C_6^r(1\text{GeV})/m_N]/a^2$     & 0.12(3) & 0.03(4) & 0.27(7) & 0.21(8)\\
\hline
degrees of freedom                    & 39-3=36 & 39-5=34 & 12-3=9 & 12-5=7 \\
$\chi^2/d.o.f.$                       & 0.4     & 0.8     & 0.1    & 1.1 \\
\hline
\hline
\end{tabular}
\end{table}

\newpage

\end{document}